\documentclass[traditabstract]{aa}
\usepackage{graphicx}
\usepackage{txfonts}
\usepackage{epsfig}
\usepackage{natbib}
\usepackage{color}
\usepackage{rotating}
\usepackage{ulem}

\newcommand{\oversim}[2]{\protect{\mbox{\lower0.5ex\vbox{%
  \baselineskip=0pt\lineskip=0.2ex
  \ialign{$\mathsurround=0pt #1\hfil##\hfil$\crcr#2\crcr\sim\crcr}}}}}
\newcommand{\simgreat}{\mbox{$\,\mathrel{\mathpalette\oversim>}\,$}} % >~ sign
\newcommand{\simless} {\mbox{$\,\mathrel{\mathpalette\oversim<}\,$}} % <~ sign
\usepackage{lineno}
\begin{document}

\title{The initial period function of late-type binary stars and its variation}
%\subtitle{The Initial Period Function}
\author{Pavel Kroupa\inst{1}, Monika G. Petr-Gotzens\inst{2}}
\institute{Argelander Institute for Astronomy, University of Bonn,
  Auf dem H\"ugel 71,D-53121 Bonn,
  Germany\\ 
\email{pavel@astro.uni-bonn.de}
\and European Southern
  Observatory, Karl-Schwarzschild-Str.\ 2, 85748 Garching,
  Germany\\ \email{mpetr@eso.org}}
\date{{\bf accepted 04.02.2011}} 

\abstract{ The variation of the period distribution function of
  late-type binaries is studied. It is shown that the Taurus--Auriga
  pre-main sequence population and the main sequence G~dwarf sample do
  not stem from the same parent period distribution with better than
  95~per cent confidence probability. The Lupus, Upper Scorpius~A and
  Taurus--Auriga populations are shown to be compatible with being
  drawn from the same initial period function (IPF), which is
  inconsistent with the main sequence data. Two possible IPF forms are
  used to find parent distributions to various permutations of the
  available data which include Upper Scorpius~B (UScB), Chameleon and
  Orion Nebula Cluster pre-main sequence samples.  All the pre-main
  sequence samples studied here are consistent with the hypothesis
  that there exists a universal IPF which is modified through
  binary-star disruption if it forms in an embedded star cluster
  leading to a general decline of the observed period function with
  increasing period.  The pre-main sequence data admit a log-normal
  IPF similar to that arrived at by Duquennoy \& Mayor (1991) for main
  sequence stars, provided the binary fraction among pre-main sequence
  stars is significantly higher.  But, for consistency with
  proto-stellar data, the possibly universal IPF ought to be flat in
  log-P or log-semi-major axis and must be similar to the K1~IPF form
  derived through inverse dynamical population synthesis, which has
  been shown to lead to the main sequence period function if most
  stars form in typical embedded clusters.

} 

\titlerunning{The initial binary-star period function} 

\authorrunning{Kroupa \& Petr-Gotzens}  

\maketitle

\section{Introduction} 
\label{sec:intro}

The initial distribution of orbital periods of binary-star systems,
i.e. the {\it initial period function} (IPF), poses an important
constraint on star-formation theory as well as being a necessary input
for modelling stellar populations.  Constraining its form, and its
possible variation with star-forming conditions, is thus of
fundamental importance. Observations of late spectral-type systems
(stellar masses $\simless 2\,M_\odot$) have shown that many young
populations have a higher binary proportion than old Galactic-field
populations (Mathieu 1994; Zinnecker \& Mathieu 2001 for reviews; and
more recently Duch\^ene et al. 2007a, 2007b, Connelley et al. 2008),
at least within a certain range of binary star periods.  Duch\^ene
(1999) has performed a comprehensive comparison of various young
stellar groups in terms of their binary properties and concluded that
very young star forming regions are very likely to have a binary
excess compared to the field or star clusters.  This conclusion is
enhanced by the study of proto-stars by Connelley et al. (2008).

However, variations between young populations also appear to be
evident, such as comparing the Taurus--Auriga (K\"ohler \& Leinert
1998) and Orion Trapezium Cluster populations (Prosser et al.\ 1994,
Petr et al.\ 1998, Petr 1998, Simon et al.\ 1999) that have a similar
age ($\approx 1$~Myr). Connelly et al. (2008) even study the possible
variation of the shape of the proto-stellar period function as
proto-stars transcend the various stages of their spectral energy
distribution evolution.  The source of such variations must be
understood in order to unearth any possible systematic variations with
the boundary conditions of their birth environment.

In this contribution we postulate the existence of a universal IPF and
test it against observations of binary period distributions for
various star forming regions.

\subsection{Binary Formation}

The formation of binary systems remains an essentially unsolved
problem theoretically (e.g. Maury et al. 2010). Fisher (2004) shows
analytically that isolated turbulent cloud cores can produce a
(unknown) fraction of binary systems with the very wide range of
orbital periods as observed. But direct cloud collapse calculations
are very limited in predicting binary-star properties owing to the
severe computational difficulties of treating the
magneto-hydrodynamics together with correct radiation transfer and
evolving atomic and molecular opacities during collapse. Available
results applying necessary computational simplifications suggest the
preference of a typical period, around\footnote{Throughout this text
  the notation $lP \equiv {\rm log}_{10}P$ is employed and $P$ is in
  days.} $lP \approx 4$, but such a narrow period distribution cannot
be transformed to the observed wide range of periods, $0\simless
lP\simless 10$ (Kroupa \& Burkert 2001).

The currently most advanced hydrodynamical simulations have been
reported by Moeckel \& Bate (2010). They allow a turbulent SPH cloud
to collapse forming a cluster of 1253~stars and brown dwarfs amounting
to $191\,M_\odot$. The cluster has a radius of about 0.05~pc and
contains a substantial binary and higher-order multiple stellar
population with a large spread of semi-major axis, $a$, but peaking at
a few~AU. After dynamical evolution with or without expulsion of the
residual gas the distribution of orbits ends up still being quite
strongly peaked at a few~AU with a significant deficit of orbits with
$a>10\,$AU when compared to the main-sequence population (their
fig.~11). This state-of-the art computation therewith confirms the
above stated issue that it remains a significant challenge for
star-formation theory to account for the Gaussian-type distribution of
$a$ spanning $10^{-1}-10^4\,$AU as observed by Duquennoy \& Mayor
(1991) and Raghavan et al. (2010) for G-dwarfs, by Mayor et al. (1992)
for K-dwarfs and by Fischer \& Marcy (1992) for M-dwarfs in the
Galactic-field.  One essential aspect which is still missing from such
computations is the stellar feedback which starts heating the cloud as
soon as the first proto stars appear. These heating sources are likely
to counter the gravitational collapse such that in reality the extreme
densities are not achieved allowing a much larger fraction of wide
binaries to survive.

More general theoretical considerations suggest that star-formation in
dense clusters ought to have a tendency towards a {\it lower} binary
proportion in warmer molecular clouds (i.e. in cluster-forming cores)
because of the reduction of available phase-space for binary-star
formation with increasing temperature (Durisen \& Sterzik 1994,
hereinafter DS).  On the other hand, an {\it enhanced} binary
proportion for orbital periods $lP \simless 5.6$ may be expected in
dense clusters due to the induction of binary formation through tidal
shear (Horton, Bate \& Bonnell 2001, hereinafter HBB), thus possibly
compensating the DS effect.  The initial period distribution function
(IPF) may thus appear similar in dense and sparse clusters, apart from
deviations at long periods due to encounters and the cluster tidal
field.  A certain fraction of stars form in small--$N$ systems, and
the dynamical decay of these is likely to affect the final
distribution of $lP\simless 5$ binaries (Sterzik \& Durisen 1998),
giving rise to non-uniform jet activity (Reipurth 2000), but again,
quantification is next-to-impossible given the neglect of the
hydrodynamical component. But, the binary formation channel must be
vastly dominating over the formation of non-hierarchical higher-order
multiples because otherwise even low-density pre-main sequence stellar
populations would have a large fraction of single stars because
initially non-hierarchical multiple systems decay on a core-crossing
time scale ($\approx 5\times 10^4\,$yr, Goodwin \& Kroupa 2005). The
low preponderance of high-order multiple proto-stars in observational
surveys is noteworthy in this context (Duch\^ene et al. 2007a).

\subsection{On the existence of a universal IPF}

Assuming binary systems emerge, by whatever process, with a universal
IPF consistent with the pre-main sequence data, and that most stars
form in embedded clusters of modest density, Kroupa (1995a,
hereinafter K1; 1995b, hereinafter K2) shows that the IPF evolves on a
few cluster crossing time-scales to the observed Galactic-field, or
present-day period function (PDPF). Applying this approach to dense
embedded clusters, Kroupa, Petr \& McCaughrean (1999, hereinafter KPM)
and Kroupa, Aarseth \& Hurley (2001, hereinafter KAH) find consistency
with the observed period function in the Orion Nebula Cluster despite
assuming the universal K1 IPF.  This suggests that observed variations
of the pre-main-sequence-star period function may be attributed to the
dynamical history of the respective population, and that there may
formally exist a universal parent IPF that emerges from the binary
star formation process and from which the observed pre-main-sequence
and main-sequence cases can be generated.

Note that this universal IPF is strictly speaking not observable due
to the rapid evolution after binary star birth, but the formal concept
of a parent IPF is useful for synthesising binary star
populations. This concept is equivalent to that of a universal IMF.
Given that the parent distribution of stellar masses which emerges
after the proto-stellar phase has been found to be invariant, it is
natural to state the {\sc Invariant Initial Binary Population
  Hypothesis} (Kroupa 2011).  Indeed, given that it is known that the
stellar IMF is invariant over a large region of physical star-forming
regime it would be expected that a universal IPF ought to exist
also. This is because the mass of a star is the last quantity to be
established in the star-formation process, with the binary property of
a stellar system being established, essentially, at a prior formation
stage.  Discarding the notion of a universal IPF, the above
star-cluster work at least shows that the dynamical history of a
population needs to be taken into account before discussing variations
of period distributions.

There are two general processes driven by fundamentally different
physics that evolve an IPF with time. On the one hand, system-internal
processes such as tidal circularisation, star-disk interactions and
decay of small$-N$ proto-systems affect binaries with periods
$lP\simless 5$. K2 referred to this collectively as {\it
  eigenevolution} of the orbital parameters. Eigenevolution leads to
period--eccentricity and period--mass-ratio correlations.  On the
other hand, binaries with $lP\simgreat 5$ are affected by encounters
in embedded clusters ({\it stimulated evolution} of the period
function).  Stimulated evolution does not significantly evolve the
eccentricity distribution of a population, but leads to a depletion of
the initial period and mass-ratio distribution function for companions
with weak binding energy (e.g. K1; K2; Parker et al. 2009).

The aim of this contribution is to quantify the significance of the
variation of the observed period distribution of various populations
of binary systems in order to allow an improved assessment of a
possible underlying parent distribution of orbital periods or
equivalently of semi-major axes. In doing so, the following questions
will be tested: {\it Is there an invariant IPF?} And, {\it which form
  does this IPF have?} Two IPF forms will be studied, namely the K1 IPF
(Eq.~\ref{eqn:fpK1} below) which had been suggested previously to
unify the pre-main-sequence and main-sequence binary populations, and
a log-normal IPF.

The IPF conceived here is a formal initial period distribution
function established after the proto-stellar phase is finished. As
such, it may not be observable, because eigenevolution affects the
shorter-period binaries on time scales of about $10^5\,$yr, while
stellar-dynamical and tidal-field influences affect it at
larger-period on the dynamical time-scale of the star-forming
system. The IPF is therefore as much a theoretical construct as the
IMF is, which is also not observable (see the {\sc Cluster IMF
  Theorem}, Kroupa 2008). The notion of a formally invariant IPF is
very powerful. If this notion is verified to be consistent with
observational data it would be important as a boundary condition for
star-formation theory (e.g. Fisher 2004), and also for the generation
of theoretical initial stellar populations such as for $N$--body
experiments of embedded clusters and 'classical' binary population
synthesis (e.g. de Kool 1996; Johnston 1996), and for dynamical
population synthesis (K2; Marks \& Kroupa 2011).

\subsection{Overview}

\noindent Section~\ref{sec:data} introduces the data sets collected
from the literature and used in this contribution for the statistical
analysis. Then, in Section~\ref{sec:ms_pms} the significance of the
difference between the observed canonical pre-main sequence and
canonical main sequence period distribution is tested. In the next
step of the present analysis (Section~\ref{sec:stfp}) all data sets
are tested for consistency with the K1 IPF and the log-normal period
function of Duquennoy \& Mayor (1991, hereinafter DM).  All data sets
are combined in Section~\ref{sec:par} to address the question if there
exists a single parent period function of the K1 or the log-normal
type, and a revised IPF is constrained using only 'matching' pre-main
sequence data in Section~\ref{sec:revfp}. The conclusions follow in
Section~\ref{sec:conc}.

%%%%%%%%%%%%%%%%%%%%%%%%%%%%%%%%%%%%%%%%%%%%%%%%%%%%%%%%%%%%%%%%%%%%%%

\section{Data}
\label{sec:data}

Brief profiles of the observational data sets used here are presented
in the sub-sections below.  In general, the binary star surveys
provide for each detected binary system the observed projected
separation, $\Delta$ (in AU), which we transform to an estimate of the
true semi-major axis, $a$, via $a = \Delta/0.95$ (Leinert et
al.\ 1993).  The values of $a$ are then converted to an orbital period
using Kepler's third law, $P_{\rm yrs}^2=a^3/M_{\rm tot}$, where
$P_{\rm yrs}=P/365.25$ is in yr and $a$ in AU, assuming the system
mass $M_{\rm tot}=1.3\,M_\odot$. This transformation is valid only
statistically and not on an individual-object basis. Note that we
choose a system mass of $1.3\,M_\odot$ to reflect the primary being of
solar mass with a typical secondary star which we take to be of about
the average stellar mass. Leinert et al.(1993) used a system mass of
$1\,M_\odot$, but this difference amounts to only about 10~per cent in
log-mass and is thus of no consequence for the present analysis.  In
the case of the DM main-sequence sample and the pre-main-sequence
binaries with $lP<4$ (data by Mathieu 1994 and Richichi et al.\ 1994),
the published $lP$ values are adopted here.

For each observational sample of $N_{\rm obs, sys}$ systems (each of
which may be a binary or a single star) the thus obtained list of $lP$
values is binned into $nb$ bins over the interval $lP_1$ to $lP_2$
covered by the observational survey (obtained from the survey interval
in~AU, $a_1$ to $a_2$, using Kepler's third law as above).  The so
constructed histogram has a bin-width $\delta{lP_i} = (lP_2-lP_1)/nb$
(in most cases equal for all bins except where noted otherwise), and
per bin $i$ of width $\delta lP_i$ there are $N_{\rm obs,bin}(lP_i)$
binaries. The observational estimate of the true period distribution
function, expressed as the binary proportion per unit $lP$ interval,
is thus obtained from
\begin{equation}
f_{\rm obs}(lP_i) = N_{\rm obs,bin}(lP_i) / (N_{\rm obs,sys}\,\delta{lP_i}), 
\label{eqn:f}
\end{equation}
with an associated binomial uncertainty (Petr et al. 1998),
\begin{equation}
e_{\rm obs}(lP_i) = \left(  {f_{\rm obs}(lP_i)\,[1 - f_{\rm obs}(lP_i)]
                    \over
                   N_{\rm obs,sys}\,\delta{lP_i} }  \right)^{1\over2}.
\label{eqn:obsbin}
\end{equation}
Binomial uncertainties are used here rather than Poisson
uncertainties, ${^{\rm P}e}_{\rm obs}(lP_i) = (f_{\rm obs}(lP_i) /
[N_{\rm obs,sys}\,\delta{lP_i}])^{1/2}$, since the typical situation
is at hand of observing $N_{\rm obs,bin}(lP_i)$ successes in $N_{\rm
  sys}$ trials, each trial having a probability $f(lP_i)$ of success,
whereby $f_{\rm obs}(lP_i)$ is an estimate of $f(lP_i)$. Note that
$e_{\rm obs}(lP_i) < {^{\rm P}e}_{\rm obs}(lP_i)$.

%.....................................................
\subsection{Main-sequence data (canonical; can.ms)}
\label{sec:ms}

The data by Duquennoy and Mayor (1992) is defined as the canonical
main sequence sample because it comprises until recently the largest,
and most detailed, multiplicity survey among low-mass main-sequence
stars. It mainly consists of G--dwarf stars, but K-- and M--dwarf
main-sequence binaries have essentially the same period distribution
(Fischer \& Marcy 1992, Mayor et al.\ 1992, and fig.~1 in K1). The
characteristics of the DM study that qualify them for the 'canonical'
main-sequence sample are the large and complete sample of solar-type
stars within 22pc of the Sun, and the fact that all periods are
covered. A key finding of the DM study is that the frequency
distribution of the binary stars' orbital periods is unimodal and
shows a well-defined log-normal distribution.  In this analysis their
published, incompleteness corrected, histogram of $lP$ values, which
is based on $N_{\rm obs,sys}=164$, is adopted.  The range in period
covered in the study is $-1.0 \le lP \le 10$ with $\delta{lP_i}=1.0$,
and $nb=11$.

A more recent survey of the multiplicity properties of solar-type
stars has been presented by Raghavan et al. (2010). This survey
comprises a sample which is about 2.5~times larger than the DM study
and reaches to larger distances. In contrast to the DM survey, it is
not based on a consistent 13-year spectroscopic survey but uses a
compendium of observations from various sources. Raghavan et
al. derive a period-distribution which they fit very well by a
log-normal form with peak at log$_{10}P = 5.03$ and standard deviation
of $\sigma($log$_{10}P) = 2.28$. The DM log-normal distribution peaks
at log$_{10} P = 4.8$ and has a standard deviation of
$\sigma($log$_{10}P) = 2.3$. The Raghavan et al. study leads therefore
to very similar results, and for the time being we continue to use the
DM survey as the canonical one.

%.....................................................
\subsection{Pre-main-sequence data (canonical, can.pms): Taurus--Auriga}
\label{sec:pms}

The pre-main sequence binary star data, i.e. the 'canonical' PMS
sample, is composed of results obtained from multiple star surveys in
the Taurus-Auriga star forming region. 

At short orbital periods, i.e.\ for $lP<2$, binaries detected by the
spectroscopic survey of Mathieu (1992, 1994) are considered.
Visual binaries in Tau-Aur, with $4.5 \le lP \le 7.5$, have been
extensively searched and studied by Ghez et al.\ (1993), Leinert et
al.\ (1993), and K\"ohler \& Leinert (1998), and the data from the
latter two publications are included in the 'canonical' PMS sample.
Slightly shorter period binaries (with $3 \le lP \le 4$) have been
addressed by Richichi et al.\ (1994), using lunar occultation
measurements. Hence, the 'canonical' PMS sample covers a significantly
large range of orbital periods, from $lP=0.5$ to $lP=7.5$, although
not continuously (see lower panel in Figure.~\ref{fig:canonfp}).  All
stars included in the 'canonical' PMS sample are low-mass stars,
similar to the stars included in the 'canonical' MS sample. The PMS
sample includes classical T~Tauri stars and weak-line T~Tauri stars
that show indistinguishable binary star period distributions (K\"ohler
\& Leinert, 1998), and are thus considered as one population in our
'canonical' PMS data.  Table~\ref{tab:logdata} lists the details for
each dataset in terms of, e.g.\ how many targets, $N_{obs,sys}$, have
been observed, which are the lower and upper survey limits (expressed
in lower and upper semi-major axis), or what distance to the region
has been assumed.

%\begin{sidewaystable*}
\begin{table*}
\begin{center}
\caption[]{Details of binary survey data used in this paper.\label{tab:logdata}}
\begin{tabular}{ccccccc}
\hline\hline\\[-3mm]
Star-forming region  & $N_{\rm obs,sys}$ & $nb$ & Lower limit & Upper limit
& Adopted distance & References \\
 & & & $a_1$ [AU] & $a_2$ [AU] & [pc] & \\

\hline

Tau-Aur (can.PMS) & 91 & 2 & lp=0.5 & lp=1.5 & 140 & (a) \\
Tau-Aur (can.PMS) & 30 & 1 & lp=3.5 & lp=3.5 & 140 & (b) \\
Tau-Aur (can.PMS) & 178 & 4 & 19.5 & 1832 & 140 & (c),(d)\\
Lupus         & 127 & 4 & 18.3 & 1348 & 160 & (e) \\
Chamaeleon assoc.   & 77  & 4 & 11.6 & 460 & 160 & (f) \\
UScA          & 68  & 3 & 18.0 & 460 & 145 & (g),(h)\\
UScB          & 51  & 3 & 18.0 & 460 & 145 & (g),(h)  \\
Orion TC      & 62  & 1 & 214  & 625 & 450 & (i),(j) \\
              & 109 & 1 & 72.9 & 214 & 450 & (i),(j) \\
\hline
\end{tabular}
\end{center}
Notes to table:
Coloumn (2): no.\ of targets surveyed. Coloumn (3): no.\ of bins 
over the survey interval [$lP_1$,$lP_2$]. Column (4): lower limit
of the survey interval in [AU], or alternatively in $lP$. Column (5): 
upper limit of the survey interval in [AU], or alternatively in $lP$.
Column (6): distance to the region. Column (7): references; (a)=
Mathieu 1994, (b)= Richichi et al.\ 1994, (c)=Leinert et al.\ 1993,
(d)=K\"ohler \& Leinert 1998, (e)=K\"ohler (priv.\ comm.), (f)=K\"ohler
2001, (g)=Brandner \& K\"ohler 1998, (h)=K\"ohler et al.\ 2000,
(i)=Petr et al.\ 1998, (j)=Petr 1998.
\end{table*}
%\end{sidewaystable*}

In the context of a possibly dynamical evolution of the binary stars'
period it is important to note that the average surface density of
young stars in Taurus--Auriga is relatively low (Gomez et
al.\ 1993). Variations or changes in the overall period distribution
due to encounters over the age of the current Taurus--Auriga PMS
population, which is $\sim$1Myr, are expected to be small. Only during
the earliest evolutionary phases of the systems, i.e. just after the
formation of the stars, is some dynamical evolution conceivable
(Kroupa \& Bouvier 2003).

%.....................................................
\subsection{Other Pre-main-sequence data}
\label{sec:pmsoth}

Beside Taurus--Aurigae, the orbital period distribution of binary
stars surveyed in the star forming regions of Lupus, Chamaeleon, Upper
Scorpius, and the Orion Trapezium Cluster (Table~\ref{tab:logdata})
are considered here. By necessity the binary fractions per separation
range as published by the respective researchers are adopted such that
here no re-analysis of possible contamination is performed as it would
go beyond the scope of this contribution.

Like for the canonical Tau-Aur PMS data, the
surveys comprise low-mass stars with a mixture of weak-line TTauri and
classical TTauri stars.  Although it cannot be ruled out completely
that some older ZAMS stars might be included in the samples, the large
majority of the surveyed targets have been confirmed to be of PMS
nature (e.g.\ Covino et al.\ 1997).  Moreover, the orbital period
distributions of the samples do not change significantly if the
samples would be restricted to confirmed PMS stars (e.g.\ K\"ohler
2001). None of the surveyed PMS populations in the analysed regions
are older than a few Myr. The Orion Trapezium Cluster, with an age of
$\sim$1\,Myr (Hillenbrandt et al.\ 1997), is likely the youngest
region, while Upper Scorpius is the oldest with $\sim$5\,Myr
(Preibisch et al.\ 2002). The star forming regions Chamaeleon and
Lupus have ages in between the Orion TC and Upper Scorpius
[Chamaeleon: 2 Myr (Luhman 2004), Lupus: $\sim$2-5 Myr (Wichmann et
  al. 1997, Makarov 2007)].
%Cha II is probably younger than Cha I (Alcala et al. 1997, AA 319, 184)

The surveyed regions do show significantly different characteristics
with respect to their average stellar densities. Nakajima et
al.\ (1998) analysed the average surface density of companions, taken
as an indicator for the strength of clustering, and find that Lupus is
the least clustered region, while the Orion TC is strongly
clustered. The stellar density in the Orion TC was found to be $\sim
5\times10^{4}$ stars/$\rm{pc}^3$ (McCaughrean \& Stauffer 1994).
Chamaeleon is considered as a loose stellar aggregate, similar to
Lupus, but with a somewhat higher average stellar density (Nakajima et
al.\ 1998). The region Upper Scorpius is part of the Sco-Cen OB
association and does not show a very high stellar density like the
Orion TC, although it is very likely that the original configuration
of Upper Sco was much denser than today, but the stars have been
dispersed with time. The stellar populations in Upper Scorpius are
typically divided into Upper Scorpius A (UScA) and Upper Scorpius B
(UScB), as these two regions show spatially distinct distributions.
Both regions are found at very similar distances (de Zeeuw et
al.\ 1999) and we assume the same overall distance value of 145\,pc as
adopted in the binary survey paper of K\"ohler et al.\ (2000). A
notably difference between UScA and UScB is that UScA contains several
high-mass (B-type) stars, while no high-mass stars are present in
UScB.  Interestingly, the observed binary period distributions in UScA
and UScB are different, with UScB showing a strong preference for
wider binaries, while in UScA mostly binaries with small separations
are present (Brandner \& K\"ohler 1998).

It should also be noted that Upper Scorpius and Orion are regions 
of low-mass {\it and} high-mass star formation, while Lupus and 
Chamaeleon are sites of low-mass star formation only.

A survey of interest in the present context is that by Ratzka et
al. (2005) and Simon et al. (1995) of $\rho$~Oph which has a density
in between that of Taurus-Auriga and the ONC. K\"ohler et al. (2000)
note that UScA and $\rho$~Oph have very similar binary distribution
functions. We will return to $\rho$~Oph and other clusters such as
Pleiades and Hyades in a more detailed investigation using Nbody
computations.

%%%%%%%%%%%%%%%%%%%%%%%%%%%%%%%%%%%%%%%%%%%%%%%%%%%%%%%%%%%%%%%%%%%%%%
\section{Main sequence versus pre-main sequence}
\label{sec:ms_pms}

In this section the statistical significance of the difference between
the observed main-sequence and the pre-main sequence binary star
period distributions is estimated. Focus is on the 'canonical' main
sequence and pre-main sequence distributions as defined in
Section~\ref{sec:data}.  The binary proportion per unit log$_{10}P$
interval is given by Eq.~\ref{eqn:f} $f(lP_i) = { N_{\rm bin}(lP_i)
  \over N_{\rm sys}\,\delta{lP_i} }$, where $N_{\rm bin}(lP_i)$ is the
number of binaries in the $i$th log-period interval $\delta{lP_i}$
($=1$ here), and $N_{\rm sys}$ is the total number of systems in the
survey, whereby each single star and each binary count as a system.
In order to obtain matching data sets, the main-sequence histogram
plotted in fig.~1 of KPM is linearly interpolated to obtain three
$lP_i$ bins with unit width for $lP_i = 5.0,6.0,7.0$ (i.e. covering
the interval $4.5 \le lP \le 7.5$). The resulting data are listed in
Table~\ref{tab:test1a2}.

\begin{table}
\begin{center}
\caption[]{Histogram of main-sequence, $f_{\rm ms}(lP_i)$, and
pre-main sequence, $f_{\rm pms}(lP_i)$, binary fractions and adopted
uncertainties.\label{tab:test1a2}}
\begin{tabular}{ccccc}
\hline\hline\\[-3mm]
$lP_i$ &$f_{\rm ms}(lP_i)$ &$e_{\rm ms}(lP_i)$   
&$f_{\rm pms}(lP_i)$ &$e_{\rm pms}(lP_i)$\\

\hline

0.5   &0.024 &0.012  &0.035 &0.038 \\
1.5   &0.049 &0.017  &0.072 &0.038 \\
3.5   &0.079 &0.022  &0.200 &0.080 \\
5.0   &0.095 &0.024  &0.164 &0.021 \\
6.0   &0.082 &0.022  &0.156 &0.021 \\
7.0   &0.076 &0.022  &0.124 &0.020 \\

\hline
\end{tabular}
\end{center}
\end{table}

The well-tried $\chi^2$ statistic is applied to test the null
hypothesis that both observed distributions stem from the same
underlying parent distribution, by calculating
\begin{equation}
\chi^2 = \sum_{i=1}^6 {\left(f_{\rm ms}(lP_i)-f_{\rm pms}(lP_i)\right)^2
         \over e_{\rm ms}^2(lP_i) + e_{\rm pms}^2(lP_i)}
\label{eqn:chi2_can}
\end{equation}
to obtain $\chi^2=15.7$ with $\nu=6$ degrees of freedom. Obtaining
such a large value, or larger, has a significance probability
$0.01<{\cal P}<0.02$, so that one can be less than 2~per cent
confident that the null hypothesis holds true. Excluding the $lP<4$
bins where the main-sequence and pre-main sequence data agree within
the uncertainties, and concentrating instead only on the more recent
Taurus--Auriga data (K\"ohler \& Leinert 1998), $\chi^2=13.2, \nu=3$
with $0.001<{\cal P}<0.01$. The null hypothesis can thus be rejected
with approximately 99~per cent confidence. It may be concluded that
the two observed distributions stem from different parent
distributions.

To obtain an additional assessment of the confidence in this result,
the Wilcoxon-Signed-Rank (WSR) test (e.g. Bhattacharyya \& Johnson
1977) provides a welcome alternative. This is a non-parametric test,
making no assumptions about the form of the underlying populations,
such as there being a well-defined mean and variance, as opposed to
using the $\chi^2$ statistic which assumes such structure in the
underlying distributions. The WSR test assesses the likelihood of
observing a certain fraction of the data asymmetrically about a
reference data set. To construct the WSR statistic, the differences
$f_{\rm pms}(lP_i)-f_{\rm ms}(lP_i)$ are ordered according to their
absolute values. These are ranked, and the ranks associated with the
positive differences are added to form the test statistic $T^+$.  The
statistic is symmetrical, that is, the same result is obtained by
considering the negative differences. For the data in
Table~\ref{tab:test1a2}, $T^+=21$ with $n=6$ data points. Obtaining
such a large or larger $T^+$ has a significance probability ${\cal
  P}=0.016$ (table~10 in Bhattacharyya \& Johnson 1977). The null
hypothesis is thus only supported with a confidence of 1.6~per cent,
confirming the above conclusion. Using only the data with $lP\ge5$
gives $n=3$ differences, which is too small for this non-parametric
test to allow significant conclusions, in contrast to the $\chi^2$
test used above, because the latter relies on additional information
about the populations.

The Kolmogorov--Smirnov test cannot be applied here nor later in this
paper, because the data sets cover different $lP$ ranges, and $f_{\rm
  P, ms}$ was estimated by DM after applying incompleteness
corrections, that is there exists no list of complete $lP$ values from
which a cumulative distribution can be generated.

%%%%%%%%%%%%%%%%%%%%%%%%%%%%%%%%%%%%%%%%%%%%%%%%%%%%%%%%%%%%%%%%%%
\section{Standard period functions} 
\label{sec:stfp}

%Differences between star forming regions exist and it is
%now an open issue if PMS populations in general show
%a differing period distribution. We are addressing this issue in the
%following by analysing the various data on binary star statistics
%of sfr now available.
Having established with a high level of confidence that the canonical
pre-main sequence and main sequence period distributions are
significantly different, one can proceed to begin inquiring as to how
the parent distributions of the two data sets may be described. In the
literature two analytic forms of possible parent distributions for the
two data sets have been used.

%...............................................
\subsection{Initial period functions (IPFs) and present-day period functions
            (PDPFs)}

For pre-main sequence binary systems K1 suggests an IPF,
\begin{equation} 
f_{\rm K}(lP) = f_{\rm pms}\,\eta \, 
                      { \left(lP - lP_{\rm min}\right) 
                      \over 
                       \delta + \left(lP - lP_{\rm min}\right)^2},
\label{eqn:fpK1}
\end{equation}
using the canonical pre-main sequence and main sequence data as
constraints. This is not in contradiction to the results of the
previous section since one takes into account through extensive
$N$body modelling that the main-sequence distribution results from the
pre-main sequence distribution if most stars form in modest (embedded)
clusters. K1 found $\eta=3.5$ and $\delta=100$ give good fits to the
pre-main sequence data, and to the main sequence data after passing
through a typical star cluster\footnote{The typical star cluster is
  the birth site of most stars in the Galaxy. According to K1 it
  contains typically 200~binaries in a characteristic radius of about
  0.8~pc, while Adams \& Myers (2001) find that most stars would
  originate from compact 10--100 member groups. This estimate can be
  argued to correspond to the previous one if residual gas loss and
  subsequent expansion with loss of stars from the modest embedded
  cluster are taken into account.} assuming the overall primordial
binary proportion is $f_{\rm pms}=1$, and setting the minimum orbital
period to 1~d ($lP_{\rm min}=0$). As seen in Fig.~\ref{fig:canonfp}
note in particular that $f_{\rm K}(lP)$ is essentially flat for
$lP\simgreat 4.5$ ($a\simgreat 20\,$AU). Eq.~\ref{eqn:fpK1} with the
above parameters is referred to as the K1 IPF, and IPFs constructed
with different values for the parameters as of the K~type.

A more elaborate model based on the above K1 IPF, but including
eigenevolution giving the correct correlation between eccentricity,
period and mass-ratio for short-period binaries, is constructed by
K2. This model has $lP_{\rm min}=1, \eta=2.5, \delta=45$ and is
required to re-produce and predict period- and mass-ratio-distribution
functions and the distribution of orbits in the eccentricity--period
diagramme for realistic stellar populations.  Such details are however
not required in the present treatment and wouldn't lead to different
results.  A useful feature of Eq.~\ref{eqn:fpK1} is that it can easily
be converted to a period-generating function (Eq.~11b in K1),
\begin{equation}
lP(X) = lP_{\rm min} +
        \left[ \delta \left( e^{2\,X\over\eta}-1 \right) \right]^{1\over2},
\label{eqn:fpK1gen}
\end{equation}
where $X\epsilon[0,1]$ is a uniform random variate, and with
$\int_{lP_{\rm min}}^{lP_{\rm max}}f_{\rm K1}(lP)\,dlP = f_{\rm pms}$.
The maximum allowed log-period, $lP_{\rm max}$, follows from
Eq.~\ref{eqn:fpK1gen} with $X=1$.  Eq.~\ref{eqn:fpK1gen} allows
efficient construction of a pre-main sequence population, such as for
$N$-body calculations of the evolution of embedded clusters (e.g. KPM;
KAH).

For the PDPF the Gaussian distribution in $lP$ describes the observed
period distribution of local G-dwarfs very well (DM),
\begin{equation}
f_{\rm DM}(lP) = f_{\rm ms}\,\kappa\,{ 1\over \sqrt{2\,\pi} \,\,
              \sigma{lP}} e^{ - {1\over2} ({lP-avlP\over
              \sigma{lP}})^2 },
\label{eqn:fpDM}
\end{equation}
where $\sigma{lP}$ and $avlP$ are, respectively, the variance in $lP$
and the average-$lP$, and $\int_{lP_{\rm min}}^{lP_{\rm max}}f_{\rm
  DM}(lP)\,dlP = f_{\rm ms}$ is enforced by adjusting $\kappa$, since
$\int_{-\infty}^{+\infty}f_{\rm DM}(lP)\,dlP = \kappa\,f_{\rm ms}$.
DM measured an overall binary proportion of $f_{\rm ms}=0.58$,
$\sigma{lP}=2.3$ and $avlP=4.8$. This is referred to as the DM PDPF,
and to period functions constructed according to Eq.~\ref{eqn:fpDM}
but with different values for the parameters as of the DM~type.  A
simple period-generating function cannot be written down; instead the
Box-Muller method for generating $lP$s is resorted to (e.g. Press et
al. 1994).

%...............................................
\subsection{The tests}
\label{sec:tests}

The aim is to test with which confidence the IPF (the K1-distribution,
Eq.~\ref{eqn:fpK1} with $\eta=3.5$, $\delta=100$ and $f_{\rm pms}=1$)
or the PDPF (the DM-distribution, Eq.~\ref{eqn:fpDM} with
$\sigma{lP}=2.3$, $avlP=4.8$ and $f_{\rm ms}=0.58$) can represent each
data set of Section~\ref{sec:data}.  For each observational sample the
theoretical distributions are converted to matching histograms by
generating $N_{\rm th,sys}=10^6$ periods and binning these into the
same $lP$-bins as in the respective observational sample, obtaining
(Eq.~\ref{eqn:f}) $f_{\rm th}(lP_i) = f_{\rm o}\,N_{\rm th,bin}(lP_i)
/ (N_{\rm th,sys}\,\delta{lP_i})$, where $f_{\rm th}(lP_i)$ is either
of the K ($f_{\rm o} = f_{\rm pms}$) or the DM ($f_{\rm o} = f_{\rm
  ms}$) type. Given the size of an observational sample, $N_{\rm
  obs,sys}$, the {\it expected} number of binaries in each $lP$ bin is
\begin{equation}
N_{\rm exp,bin}(lP_i) = f_{\rm th}(lP_i)\, N_{\rm obs,sys}\, \delta{lP_i}
\label{eqn:Nexp}
\end{equation}
with associated expected binomial uncertainty 
\begin{equation}
e_{\rm N,exp,bin}(lP_i) = \left[  f_{\rm th}(lP_i)\,(1 - f_{\rm th}(lP_i))\,
                   \delta{lP_i}\,N_{\rm obs,sys}  \right]^{1\over2}.
\label{eqn:thbin}
\end{equation}
To quantify the goodness-of-fit the $\chi^2$ statistic for the 
K or DM distribution becomes,
\begin{equation}
\chi_{\rm K}^2 ({\rm or} \chi_{\rm DM}^2) = \sum_{i=1}^{nb} \left(
    {N_{\rm exp,bin}(lP_i) - N_{\rm obs,bin}(lP_i) \over e_{\rm
        N,exp,bin}(lP_i)} \right)^2,
\label{eqn:chi2_mod}
\end{equation}
with $nb$ degrees of freedom since no free parameters are
fitted. Terms with $N_{\rm exp,bin}(lP_i)=0$ give infinite $\chi^2$
since a finite datum is inconsistent with the model, which is treated
as having no intrinsic uncertainties.  The significance probability,
${\cal P}$, of obtaining $\chi^2 \ge \chi_{\rm K}^2 ({\rm or}
\chi_{\rm DM}^2)$, is evaluated using the incomplete gamma function
(Press et al. 1994).

%...............................................
\subsection{Results}
\label{sec:results}

\begin{figure}
\includegraphics[angle=0,scale=0.43]{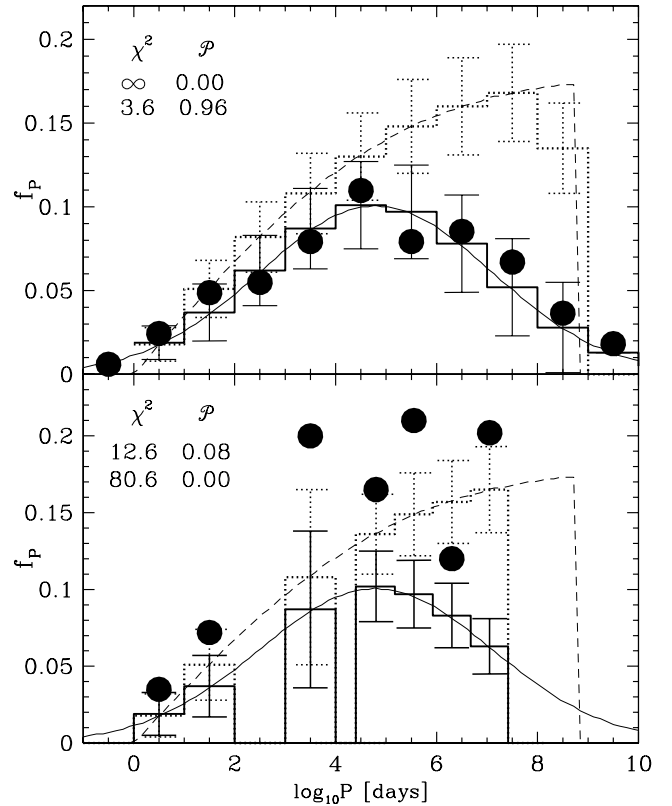}
\vskip 0mm
\caption{Testing the likelihood that the observed canonical binary-star
    period distributions fit the K1 IPF (dashed curve) or the DM PDPF
    (solid line). The model histograms are shown as dotted and solid
    lines with expected binomial uncertainties as error-bars
    (Eqs~\ref{eqn:Nexp} and~\ref{eqn:thbin}). Solid circles are the
    G-dwarf main sequence data in the upper panel and canonical
    pre-main sequence data in the lower panel
    (Section~\ref{sec:data}).  The $\chi^2$ value and significance
    probability, ${\cal P}$, of observing such a large or larger
    $\chi^2$ is written in each panel (upper numbers for testing the
    data against the K1 IPF, lower numbers for testing against the DM
    PDPF.)  }
\label{fig:canonfp}
\end{figure}

The above procedure is applied to all the data and the results are
shown in Fig.~\ref{fig:canonfp}. The conclusion with very high
confidence (at the 99.5~per cent level or better) is that the
canonical main-sequence data are not consistent with the K1 IPF (upper
panel), and that the canonical pre-main sequence data are not
consistent with the DM PDPF (lower panel). This does not change even
if the $lP=9.5$ main-sequence value is ignored in the upper panel
(giving $\chi^2=43$ with ${\cal P}=0.00$ for the K1 IPF). This result
is consistent with that of Section~\ref{sec:data} where it was shown
that the canonical main-sequence and pre-main-sequence data are
significantly different.

%======================================================
% ~/PAPERS/Petrpap/ProgsaFigs/canonper.sm
%====================================================

\begin{sidewaysfigure*}
\begin{center}
\includegraphics[angle=-90,scale=0.8]{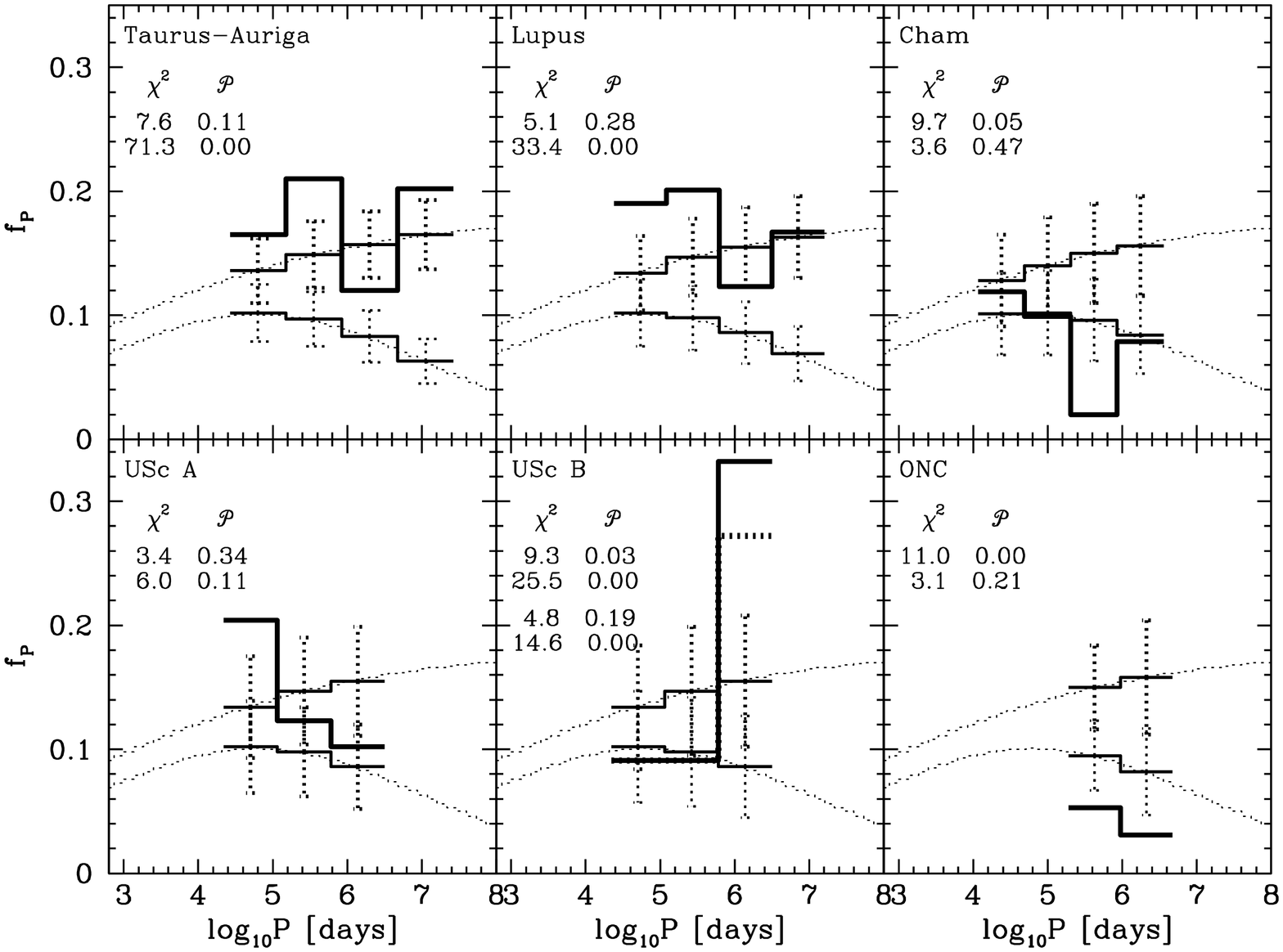}
%\vskip -10mm
\caption
{Similar to Fig.~\ref{fig:canonfp} but for all pre-main
sequence samples.  In all panels the dotted continuous curves are the
model distributions (K1 IPF upper curves, DM PDPF lower curves),
whereas the model histograms are shown as solid lines with expected
binomial uncertainties as dotted error-bars. Thick histograms are the
data (Section~\ref{sec:data}).  The $\chi^2$ value and probability,
${\cal P}$, of observing such a large or larger $\chi^2$ is written in
each panel (upper numbers for testing the data against the K1 IPF,
lower numbers for testing against the DM IPF).  For UScB, the lower
set of numbers refers to the dotted histogram, which results after
removing the three widest binaries from the sample.}
\label{fig:allfp}
\end{center}
\end{sidewaysfigure*}

The results of applying this procedure to each of the pre-main
sequence data sets are presented in Fig.~\ref{fig:allfp}.  
%======================================================
% ~/PAPERS/Petrpap/ProgsaFigs/allper.sm
%====================================================
It is evident that the Taurus--Auriga, Lupus and UScA populations are
consistent with the hypothesis that they be drawn from the K1 IPF. The
Cham pre-main sequence population is marginally consistent with this
hypothesis that the parent distribution is the K1 IPF at the 5~per
cent confidence level, whereas the full UScB sample is only consistent
with the K1 IPF at the 3~per cent confidence level.  UScB becomes
consistent at the 19 per cent confidence level with the K1 IPF if
three of the widest binaries are removed from the sample, assuming
these are significantly closer ($<90$\,pc) than the bulk of UScB stars
at $\sim 145$\,pc.  Their apparent brightness is clearly brighter than
expected for their spectral type when placed at 145\,pc, and proper
motion measurements confirm nearby distances for at least two of the
wide binary systems (Perryman et al.\ 1997; Salim \& Gould 2003).

On the other hand, an alternative hypothesis is that the data sets be
drawn from the DM PDPF. This hypothesis can be discarded with very
high confidence for the Taurus--Auriga, Lupus and UScB pre-main
sequence populations. It cannot be discarded for the Cham, UScA and
ONC populations. The Cham and UScA populations can, in fact, be drawn
from either period function. The enhanced significance probability for
the hypothesis that Cham be drawn from the DM~PDPF nevertheless allows
the hypothesis that Cham started off in a clustered mode with a
K1~IPF, the Cham population possibly being an evolved version of
the canonical pre-main sequence population. This also holds true for
the ONC sample as shown by KPM and KAH.

%%%%%%%%%%%%%%%%%%%%%%%%%%%%%%%%%%%%%%%%%%%%%%%%%%%%%%%%%%%%%%%%%%%%%%
\section{Parent period distribution functions} 
\label{sec:par}

In Section~\ref{sec:results} the K1~IPF and the DM~PDPF were tested
against all the available data sets individually. 

\begin{figure}
\includegraphics[angle=0,scale=0.43]{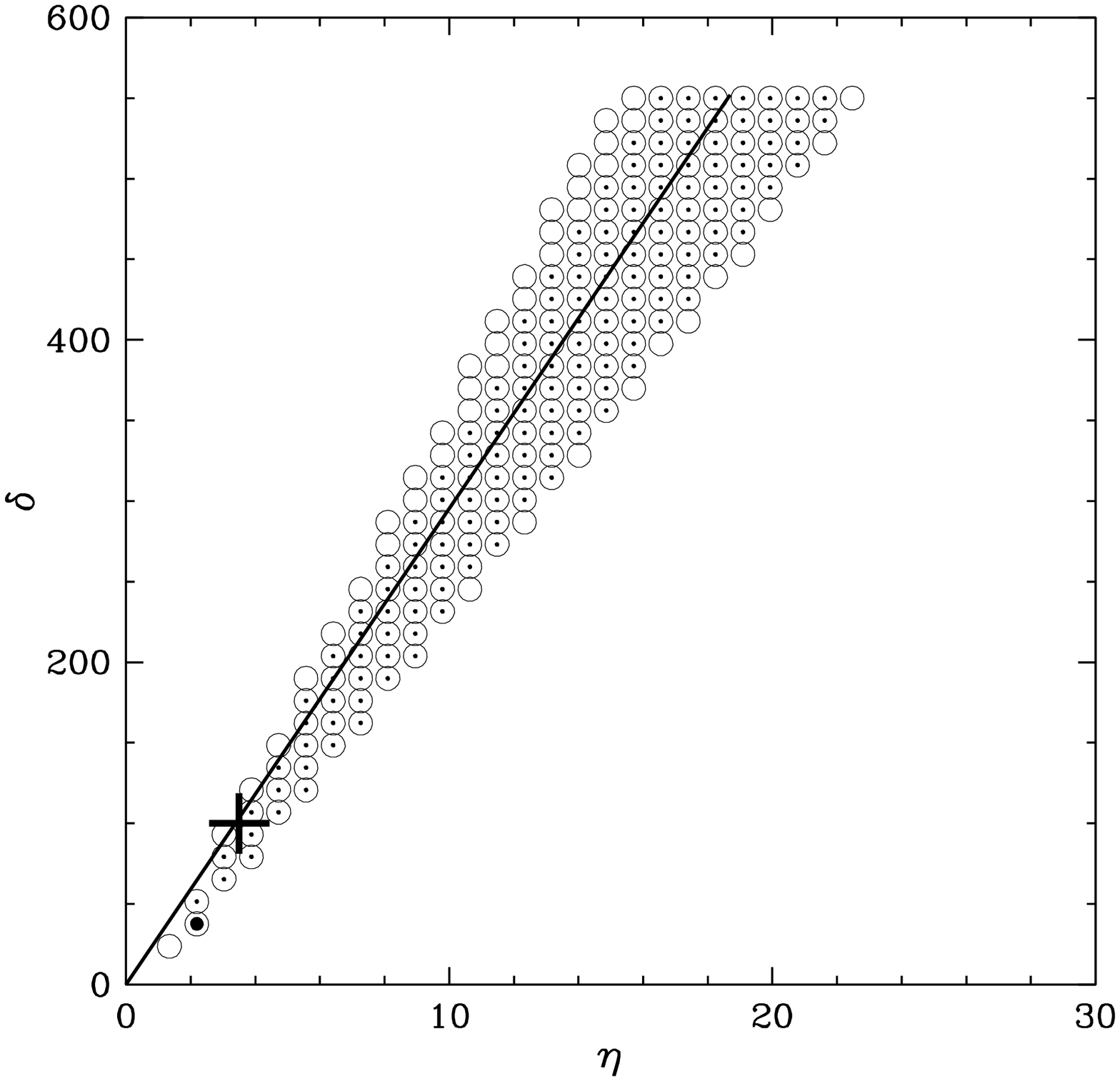}
\vskip -20mm
\caption{Survey of $\eta,\delta$ parameter space for solutions of the
    K~type with $f_{\rm pms}=1$ (Eq.~\ref{eqn:fpK1}) using the
    canonical pre-main sequence data. Open circles and small dots
    delineate models consistent with the data at the 1 and 5~per cent
    confidence level or better, respectively, whereas the thick solid dot
    shows a model consistent at the 50~per cent level or better. The
    K1~IPF is shown as the thick cross. Note that the K1~IPF is not
    the same as the best solution found here. This is not surprising
    because the K1~IPF was derived by constraining the IPF by both the
    pre-main sequence and main-sequence period distribution
    functions. The best solution found here thus lacks one major
    constraint which lead to the K1~IPF, but both are sufficiently
    close to conclude that the K1~IPF is a good solution to the
    pre-main sequence data alone.  The straight line shows the
    asymptotic solution for large $\delta$: $f_{\rm K1}(lP) =
    (\eta/\delta)\,(lP-lP_{\rm min})$, with $\eta/\delta=0.0339$.}
\label{fig:sK1}
\end{figure}

\begin{figure}
\includegraphics[angle=0,scale=0.43]{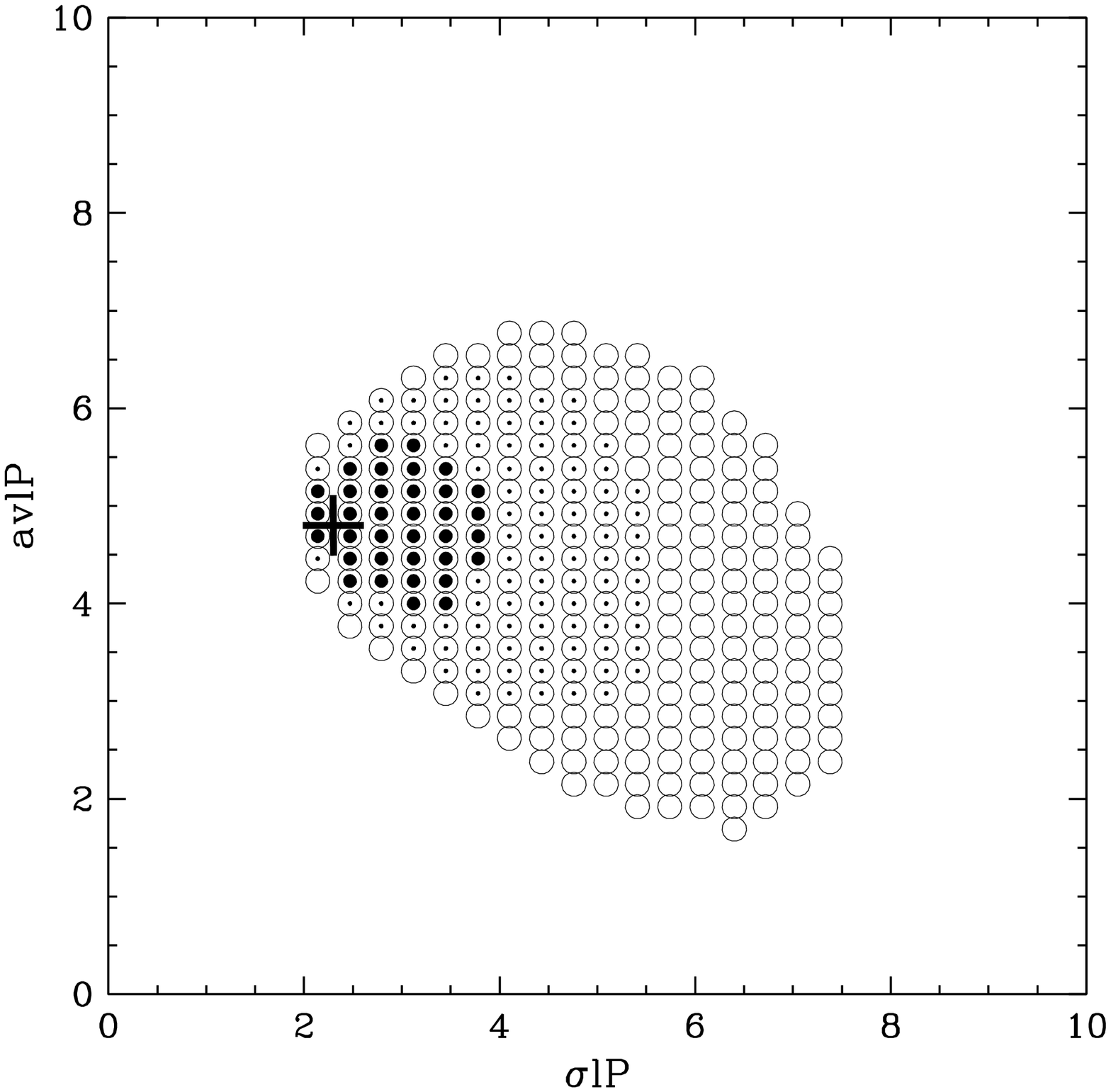}
\vskip -20mm
\caption{Survey of $\sigma{lP},avlP$ parameter space for solutions of
the DM~type with $f_{\rm ms}=0.58$ (Eq.~\ref{eqn:fpDM}) using the
canonical main sequence data. Open circles and small dots delineate
models consistent with the data at the 1 and 5~per cent confidence
level or better, respectively, whereas thick solid dots show models
consistent at the 50~per cent level or better. The DM~PDPF is shown as
the thick cross.}
\label{fig:sG}
\end{figure}

In this section data sets are combined and more general solutions for
possible parent distributions of the K and DM~types
(Eqs~\ref{eqn:fpK1} and~\ref{eqn:fpDM} respectively) are sought by
scanning the parameter spaces $\eta,\delta$ and $\sigma{lP},avlP$
using suitable increments by evaluating Eq.~\ref{eqn:chi2_mod} and the
associated significance probability ${\cal P}$. This procedure is
demonstrated in Figs.~\ref{fig:sK1} and~\ref{fig:sG} by first of all
finding all admittable solutions to the canonical data sets. The
solutions so found agree with the known solutions (the K1~IPF and
DM~PDPF), and one may be confident that the algorithm functions.
%======================================================
% ~/PAPERS/Petrpap/ProgsaFigs/surv_K1fp.sm
%====================================================
%======================================================
% ~/PAPERS/Petrpap/ProgsaFigs/surv_Gauss.sm
%====================================================

On combining $n_{\rm dat}$ data sets under a common hypothesis, the
combined $\chi_{\rm comb}^2$ and number of degrees of freedom, $nb_{\rm
comb}$, are computed,
\begin{eqnarray}
nb_{\rm comb} = \sum_i^{n_{\rm dat}} nb_i, \nonumber\\
\chi_{\rm comb}^2 = \sum_i^{n_{\rm dat}} \chi_i^2,
\label{eqn:combchi}
\end{eqnarray}
where $\chi_i^2$ and $nb_i$ are the chi-square and number of degrees
of freedom of data set $i$.  Possible data sets are $i =$ Lupus, Cham,
UScA, UScB, ONC, canonical pre-main sequence (=can.pms), canonical
main sequence (=can.ms)$]$.  The significance probability, ${\cal
P}_{\rm KorDM}$, of obtaining $\chi^2 \ge \chi_{\rm comb}^2$, is
evaluated as above (Eq.~\ref{eqn:chi2_mod}), and ${\cal P}_{\rm K},
{\cal P}_{\rm DM}$ are the resulting combined confidence probabilities
on using the K~and DM~type PFs, respectively, for testing the
hypotheses set up in the following. A hypothesis is deemed consistent
with the data if ${\cal P}_{\rm KorDM} \ge 5\times10^{-2}$, i.e. if
the significance probability is 5~per cent or better.

The following hypotheses are tested:

\begin{enumerate}

\item {\it A PF of the K~type is the parent distribution of all
combined data sets} ($n_{\rm dat}=7$).

Assuming $f_{\rm pms}=1$ and scanning the parameter space, as above,
yields no solutions at all in agreement with the data at the 0.001~per
cent level or better (${\cal P}_{\rm K}\le 10^{-5}$). Thus the
hypothesis can be discarded at the 99.999~per cent level. Assuming
$f_{\rm pms}=0.8$ and $f_{\rm ms}=0.6$ also leads to rejection of this
hypothesis with ${\cal P}>99.999$~per cent confidence.

The general conclusion, stated with a confidence better than 99.999~per
cent, is thus that there is no single parent distribution of the
K~type for all data sets combined.

\item {\it A PF of the DM~type is the parent distribution of all
combined data sets} ($n_{\rm dat}=7$).

Assuming $f_{\rm ms}=0.58, 0.8, 1.0$ and scanning the parameter space,
as above, yields no solutions at all in agreement with the data at the
0.001~per cent level or better (${\cal P}_{\rm DM}\le 10^{-5}$).

Thus there is no single Gaussian parent distribution of all data sets
combined. The hypothesis is rejected with a confidence better than
99.999~per cent. 

\item {\it A PF of the K~type is the parent distribution of the
combined canonical pre-main sequence and canonical main sequence data
sets} ($n_{\rm dat}=2$).

The relevant confidence is determined from Eq.~\ref{eqn:combchi} with
[$i=$can.pms, can.ms]. The result is that there is no solution with
confidence better than ${\cal P}_{\rm K}=10^{-5}$ for $f_{\rm
  pms}=1,0.8,0.6$, so that this hypothesis can be discarded with a
confidence better than 99.999~per cent.

\item {\it A PF of the DM~type is the parent distribution of the
combined canonical pre-main sequence and canonical main sequence data
sets} ($n_{\rm dat}=2$).

The relevant confidence is determined from Eq.~\ref{eqn:combchi} with
[$i=$can.pms, can.ms]. There is no solution with confidence better
than ${\cal P}_{\rm DM}=10^{-5}$ for $f_{\rm ms}=0.6$. For $f_{\rm
  ms}=0.8$ the parameter region $\sigma{lP} = 2.5\pm0.4,
avlP=5.1\pm0.2$ has ${\cal P}\le 0.1$~per cent, while for $f_{\rm
  ms}=1.0$ the parameter region $\sigma{lP} = 2.6\pm0.4,
avlP=5.2\pm0.2$ has ${\cal P}\le 0.05$~per cent.  This hypothesis can
thus be discarded with 99.9~per cent confidence.

\item \label{hyp5} {\it A PF of the K~type is the simultaneous parent
distribution of the canonical pre-main sequence data, the Lupus and
UScA pre-main sequence data sets} ($n_{\rm dat}=3$).

The relevant confidence is determined from Eq.~\ref{eqn:combchi} with
[$i=$can.pms, Lupus, UScA]. These are chosen because
Fig.~\ref{fig:allfp} indicates these data sets to be similar in that
they appear more or less unevolved according to the results of K1 and
K2. Scanning $\eta,\delta$ parameter space leads to solutions with
${\cal P}_{\rm K} > 10^{-2}$ if $f_{\rm pms}=1.0, 0.8$
(Fig.~\ref{fig:sK_c1} and~\ref{fig:sK_c2}, respectively), whereas
${\cal P}_{\rm K} < 10^{-5}$ if $f_{\rm pms}=0.6$. Thus, a
simultaneous parent distribution is only possible if the pre-main
sequence binary-star fraction is larger than the main sequence value
of~60~per cent.
%======================================================
% ~/PAPERS/Petrpap/ProgsaFigs/check_surv_K1fp.sm
% with p = p5 and using scan_1_22.dat (fpms=1.0)
\begin{figure}
\includegraphics[angle=0,scale=0.43]{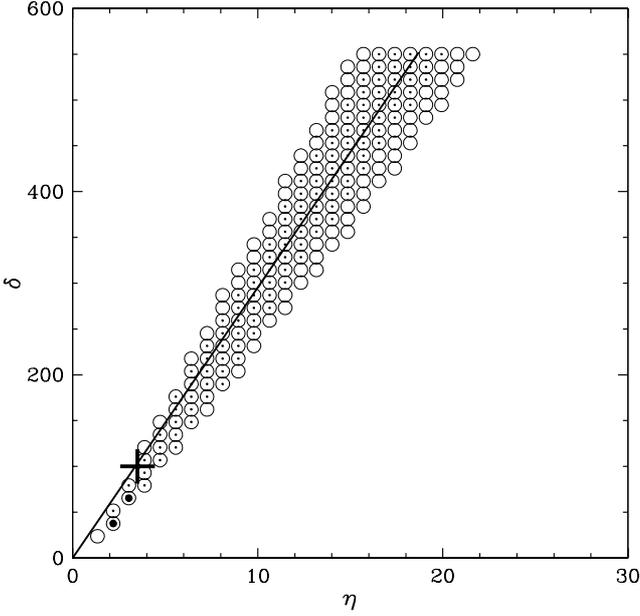}
\vskip -20mm
\caption
{Testing hypothesis~\ref{hyp5} with $f_{\rm pms}=1.0$.  Survey
    of $\eta,\delta$ parameter space for solutions of the K~type with
    $f_{\rm pms}=1$ (Eq.~\ref{eqn:fpK1}) to the combined canonical
    pre-main sequence data, Lupus and UScA pre-main sequence data
    sets.  Open circles and small dots delineate models consistent
    with the data at the 1 and 5~per cent confidence level,
    respectively. The K1~IPF is shown as the thick cross, and the
    straight line is as in Fig.~\ref{fig:sK1}.  The thick solid dots
    show models consistent at the 50~per cent level or better.}
\label{fig:sK_c1}
\end{figure}
%====================================================
%======================================================
% ~/PAPERS/Petrpap/ProgsaFigs/check_surv_K1fp.sm
% with p = p5 and using scan_2_22.dat (fpms=0.8)
\begin{figure}
\includegraphics[angle=0,scale=0.43]{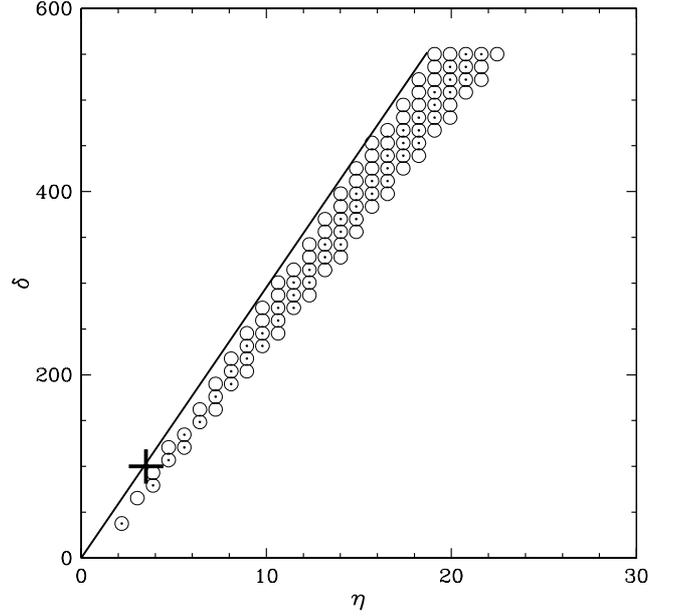}
\vskip -20mm
\caption{As Fig.~\ref{fig:sK_c1}. 
Testing hypothesis~\ref{hyp5} with $f_{\rm pms}=0.8$.}
\label{fig:sK_c2}
\end{figure}
%====================================================

\item \label{hyp6} {\it A PF of the DM~type is the simultaneous parent
distribution of the canonical pre-main sequence data, the Lupus and
UScA pre-main sequence data sets} ($n_{\rm dat}=3$). 

Scanning $\sigma{lP},avlP$ parameter space leads to solutions with
${\cal P}_{\rm DM} > 10^{-2}$ if $f_{\rm ms}=1.0$
(Fig.~\ref{fig:sDM_c}), whereas ${\cal P}_{\rm DM}< 5\times10^{-4}$ if
$f_{\rm ms}=0.8$ and ${\cal P}_{\rm DM} < 10^{-5}$ if $f_{\rm
ms}=0.6$.  Thus, a simultaneous Gaussian parent distribution is only
possible, with a significance probability better than 5~per cent, if
the pre-main sequence binary-star fraction is significantly larger
than the main sequence value of~60~per cent (note that $f_{\rm ms}$ in
Eq.~\ref{eqn:fpDM} is here the binary-star fraction required for the
combined pre-main sequence samples).
%======================================================
% ~/PAPERS/Petrpap/ProgsaFigs/check_surv_Gauss.sm
% with p = p6 and using scan_3_22.dat (fms=1.0)
\begin{figure}
\includegraphics[angle=0,scale=0.43]{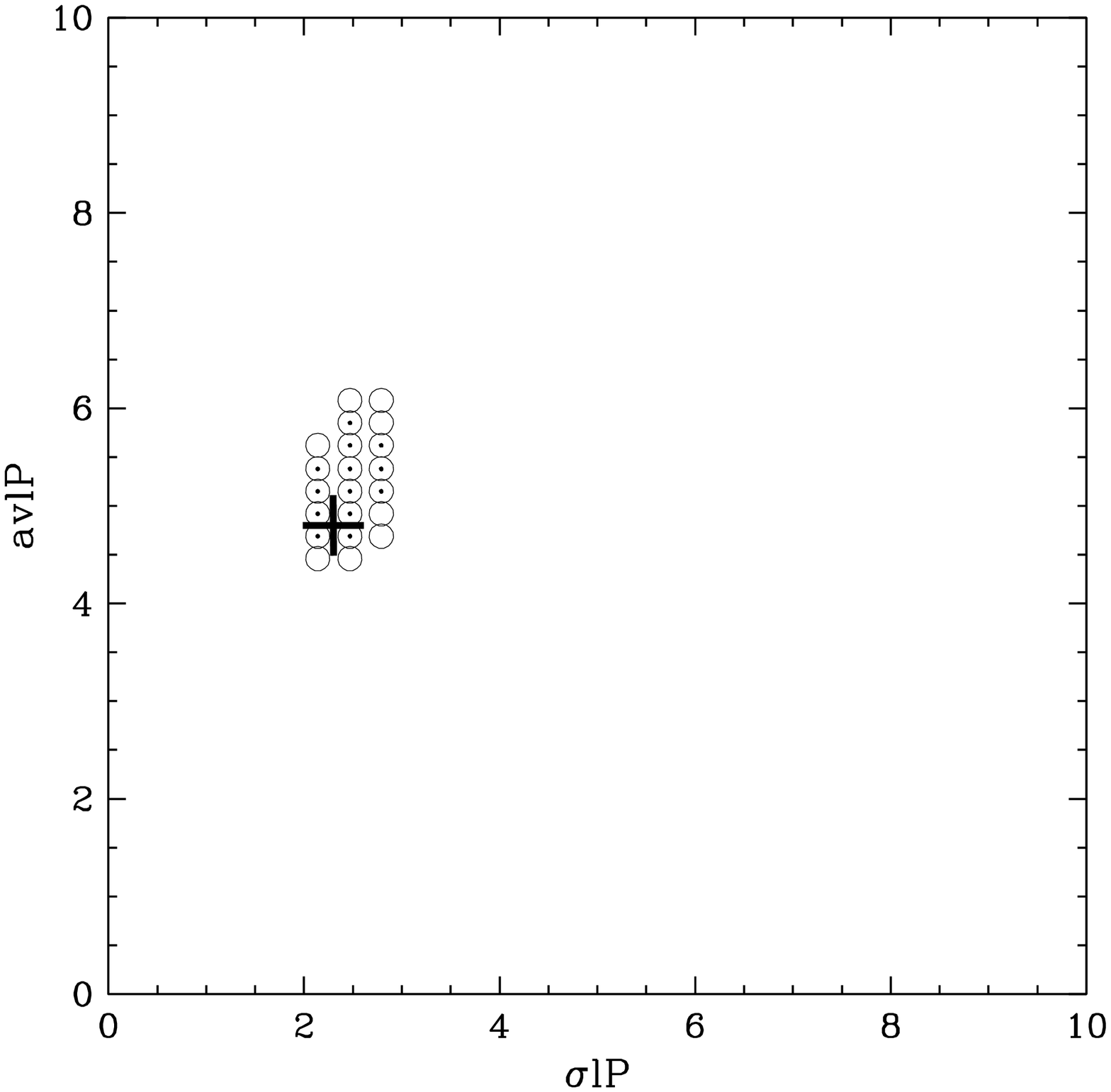}
\vskip -20mm
\caption
{Testing hypothesis~\ref{hyp6} with $f_{\rm ms}=1.0$.  Survey of
$\sigma{lP},avlP$ parameter space for solutions of the DM~type
(Eq.~\ref{eqn:fpDM}) to the combined canonical pre-main sequence data,
Lupus and UScA pre-main sequence data sets.  Open circles and small
dots delineate models consistent with the data at the 1 and 5~per cent
confidence level, respectively. The DM~PDPF is shown as the thick
cross. }
\label{fig:sDM_c}
\end{figure}
%====================================================

\item {\it A PF of the K~type is consistent with all pre-main sequence
data (canonical, Lupus, UScA, Cham, UScB and ONC)} ($n_{\rm dat}=6$).

No parent distribution of the K-type can be found with 0.1~per cent
confidence or better. This suggests that these pre-main sequence data
sets may either have different IPFs, or that they stem from the same
IPF but were modified, as would be expected for example for the ONC
data (KPM, KAH).  UScB, however, forms a definite outlier, since it
has a significant surplus of binary stars at long periods
(Fig.~\ref{fig:allfp}), when taking the observational results at face
value as we do here for all datasets. After removing possible
non-members (cf.\ Sec.~\ref{sec:results}), a surplus of binary stars
at long periods persists although less significant, which cannot be
understood through binary-star disruption in embedded clusters.

\item {\it A PF of the~DM~type is consistent with all pre-main
sequence data (canonical, Lupus, UScA, Cham, UScB and ONC)}
($n_{\rm dat}=6$).

No DM~type PF exists with ${\cal P}_{\rm DM}>5\times10^{-4}$, so that
this hypothesis can be rejected with 99.95~per cent confidence.

\end{enumerate}

%%%%%%%%%%%%%%%%%%%%%%%%%%%%%%%%%%%%%%%%%%%%%%%%%%%%%%%%%%%%%%%%%%%%%%
\section{A revised IPF} 
\label{sec:revfp}

In the previous section, the tests of hypotheses~\ref{hyp5}
and~\ref{hyp6} yield parameter ranges for Eqs~\ref{eqn:fpK1}
and~\ref{eqn:fpDM} that are consistent with the combined data at the
5~per cent confidence level or better. These parameter ranges are
shown in Figs~\ref{fig:sK_c1} and \ref{fig:sK_c2} for K-type
solutions, and in Fig.~\ref{fig:sDM_c} for DM-type solutions. The data
sets were selected according to appearing least evolved, using the
a-priori knowledge gained in K1 and K2.

It is noteworthy that the K-type solutions ($\eta \approx 2.5-3,
\delta\approx 40-70$) are consistent within 5~\% confidence with the
K1~IPF ($\eta=3.5, \delta=100$), thus corroborating the results
obtained in K1. This is particularly interesting because here we used
data sets including more than one pre-main sequence population, while
in K1 only older Taurus--Auriga data were available but no functional
fitting to these data was performed such as here. Rather, the K1~IPF
was obtained through inverse dynamical population synthesis by
demanding approximate (eye-ball) fits to the pre-main sequence, and to
the main sequence data after star-cluster disintegration. The present
purely statistical approach confirms the K1~IPF ($\eta\approx3.5,
\delta\approx 100, f_{\rm pms}\ge 0.8$, with enhanced confidence
probability for larger $f_{\rm pms}$).

A new result obtained here (Fig.~\ref{fig:sDM_c}) is that a log-normal
function in orbital period $P$ can also be considered a parent
distribution of the same pre-main sequence data as above. These data
are consistent, at the 5~per cent confidence level or better, with
being drawn from a parent distribution of the DM-type, if this PF has
parameters indistinguishable from the DM~PDPF ($\sigma{lP}\approx2.3,
avlP\approx5$) but with a significantly larger binary proportion,
$f_{\rm ms}=1.0$.

\begin{figure*}
%\rotatebox{-90}{\resizebox{0.7 \textwidth}{!}
\centering
\includegraphics[angle=-90,scale=0.7]{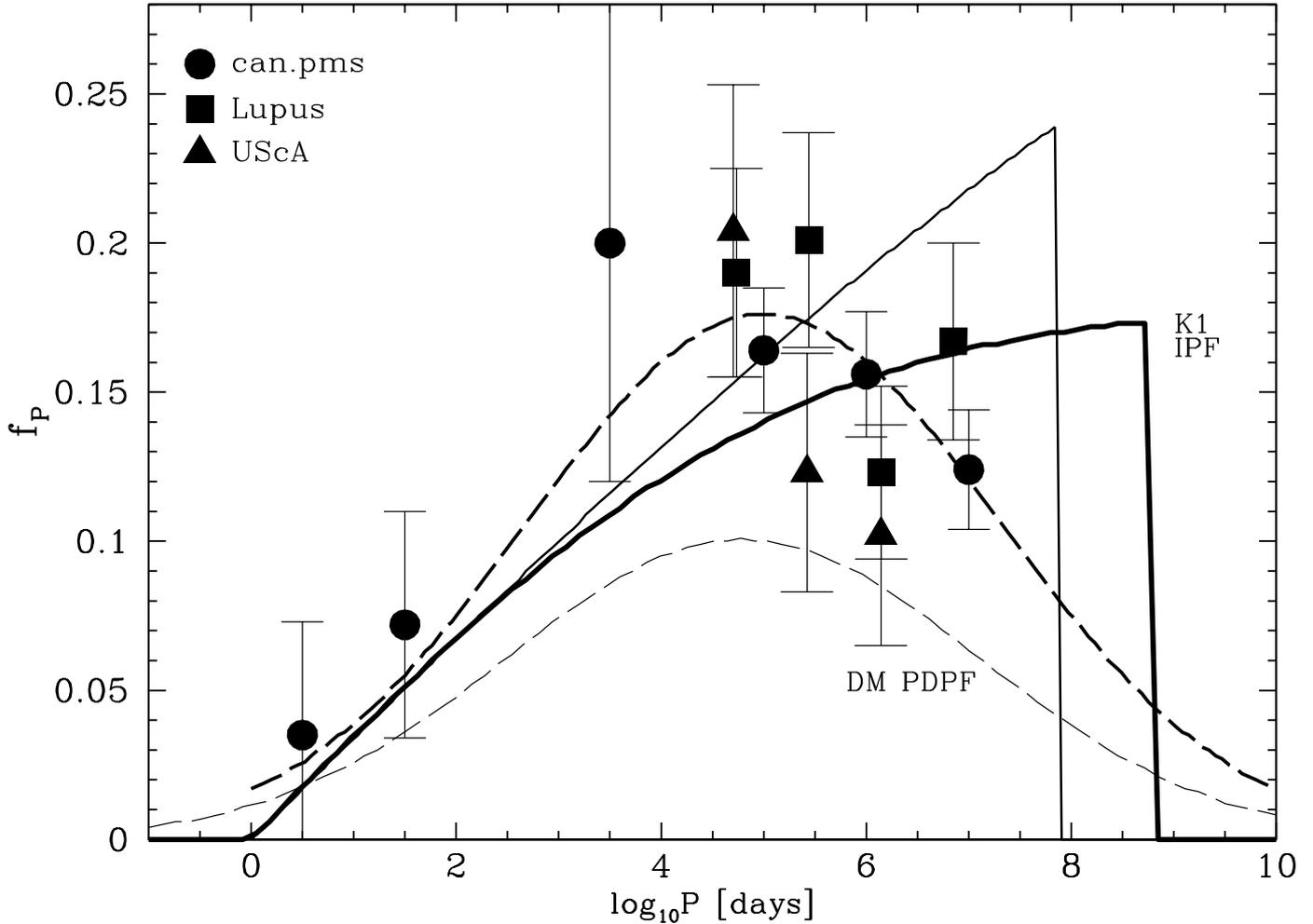}
\vskip 0mm
\caption
{The revised IPF. In addition to the K1~IPF (thick solid
curve), two possible IPF forms are displayed. These are identified as
solutions in Figs.~\ref{fig:sK_c1} and~\ref{fig:sDM_c}, and are shown
as the thin solid line (K-type, Eq.~\ref{eqn:fpK1} with the asymptotic
solution $\delta = 29.52\,\eta; \eta=18.7$ and $f_{\rm pms}=1.0$) and
the thick dashed line (DM-type, Eq.~\ref{eqn:fpDM} with
$\sigma{lP}=2.3, avlP=5.0, f_{\rm ms}=1.0$). The thin dashed line is
the DM~PDPF. Data points with observational error bars are as
indicated in the key.  These are used to test hypothesis ~\ref{hyp5}
and~\ref{hyp6} giving ${\cal P}=0.10$ (thin solid curve), ${\cal
P}=0.22$ (thick dashed curve) and ${\cal P}=0.10$ (thick solid curve).}
\label{fig:revIPF}
\end{figure*}

The result is thus that either the K1~IPF (Eq.~\ref{eqn:fpK1}), or a
K-type solution with $\delta \approx 29.52\,\eta$ (straight line in
Fig.~\ref{fig:sK_c1}), or the Gaussian PF (Eq.~\ref{eqn:fpDM}) with
the above parameters can be parent distributions of pre-main sequence
binary systems with indistinguishable confidence probabilities. This
is illustrated in Fig.~\ref{fig:revIPF}.
%======================================================
% ~/PAPERS/Petrpap/ProgsaFigs/revIPF.sm
%====================================================

%%%%%%%%%%%%%%%%%%%%%%%%%%%%%%%%%%%%%%%%%%%%%%%%%%%%%%%%%%%%%%%%%%%%%%
\section{Conclusions} 
\label{sec:conc}

A fresh look is taken at the variations evident in the period
distribution of binary stars in various stellar populations. The most
notable such difference, noted by many authors, lies between the
canonical pre-main sequence population of Taurus--Auriga and the
canonical Galactic-field local G-dwarf population. The difference is
significant (Section~\ref{sec:ms_pms}), but can be understood to
result from stimulated evolution of an initial Taurus--Auriga-type
population if most Galactic-field stars originate in modest embedded
clusters (K1).  That work arrives at a possible initial period
function, the K1~IPF (Eq.~\ref{eqn:fpK1}), which is nearly flat for
$a\simgreat 20\,$AU.

The variation between pre-main sequence populations is also studied
here, with the result that significant differences are evident,
notably between the canonical pre-main sequence population and the ONC
and UScB samples.  Stimulated evolution in the dense Orion Nebula
Cluster efficiently depletes a K1~IPF to the observed data (KPM, KAH),
so that the former difference can be readily accounted for. The latter
difference, which results from a significant apparent surplus of
long-period binaries may indicate that the IPF of the UScB population
was very different to that in Taurus--Auriga, as noted by Brandner \&
K\"ohler (1998). Noteworthy in this context is that UScA, where mostly
binaries with small separations are present, contains several
high-mass (B-type) stars, while no high-mass stars are present in
UScB, which shows a preference for wider binaries. This is
qualitatively consistent with UScA perhaps being the result of denser
embedded stellar groups that are already dispersed after dynamical
evolution and the expulsion of residual gas by the massive
stars. Discarding three of the widest binaries in the UScB data set
yields a binary population consistent with the K1 IPF
(Sect.~\ref{sec:results}), leaving the ONC as the only pre-main
sequence population being clearly different from the canonical
pre-main sequence population. In any case, among the data sets used in
this paper the UScB data has the lowest number of systems surveyed for
binarity, and the region is most likely more affected by containing
dispersed mixed populations of stars with different distances and ages
than the other data. Hence, the observed orbital period distribution
of binaries in UScB remains to be confirmed.

The (canonical) Taurus--Auriga, Lupus and UScA populations appear
least-evolved and are consistent with being drawn from a common parent
IPF {\it iff} the binary proportion is higher than on the main
sequence. The set of possible IPFs is shown in Figs.~\ref{fig:sK_c1},
\ref{fig:sK_c2} and~\ref{fig:sDM_c}, noting that the K1~IPF is
included. Three possible IPFs, all with indistinguishable confidence
probabilities, are presented in Fig.~\ref{fig:revIPF}.  While no
formal decision can be made based on ${\cal P}_{\rm KorDM}$ as to
which of the three IPFs plotted in Fig.~\ref{fig:revIPF} are to be
preferred, the a-priori knowledge gained in K1 favours the K1~IPF,
because it is~(1) consistent with the data at the 10~per cent
confidence level, (2)~it is the precursor of the Galactic-field PDPF
if most stars form in embedded clusters, and~(3) the revised K-type
IPF shown as the thin solid line in Fig.~\ref{fig:revIPF} leads to a
deficit of Galactic-field binaries with $lP\simgreat 6$ (fig.~8 in
K1).  Inverse dynamical population synthesis will have to be applied
to investigate if the alternative log-normal IPF (thick dashed line in
Fig.~\ref{fig:revIPF}) can be made consistent with the Galactic-field
PDPF for a reasonable library of embedded star clusters. However, by
consulting fig.~1 of Connelley et al. (2008) it becomes readily
apparent that a flat distribution function for $a\simgreat 100\,$AU is
preferred over a log-normal function, as is also concluded by those
authors and as is the case for the K1~IPF.  This issue is studied in
more detail now (Marks, Kroupa \& Oh 2011) and does suggest that the
K1~IPF (Eq.~\ref{eqn:fpK1}) is favoured. Parker et al. (2009) come to
a similar conclusion and stress that $a>10^4\,$AU binaries cannot
survive in any clustered environment. They may however form during
cluster dissolution (Kouwenhoven et al. 2010).

It is notable that the above three pre-main sequence populations, and
in addition the Cham and ONC populations, all are consistent with a
monotonically decreasing PF with increasing $lP$
(Figs.~\ref{fig:allfp} and~\ref{fig:revIPF}). This may suggest that
all known pre-main sequence populations have already suffered some
degree of stimulated evolution, and may have begun with the same
(universal) IPF. This notion allows reconstruction of the properties
of the embedded clusters from which the pre-main sequence populations
could have originated (inverse dynamical population synthesis).
Explicit modelling of this evolution has been performed for the ONC
(KPM, KAH; Parker et al.2009) and sparse embedded clusters (K1, K2,
Kroupa \& Bouvier 2003; Parker et al. 2009).

Thus, at the present it cannot be confirmed that the IPF varies with
star-forming conditions. That is, the presently available data are
consistent with an invariant birth or initial binary population.

\bibliographystyle{aa}

\end{document}